\documentclass[12pt]{article}
\usepackage{color}
\usepackage{amsmath,amssymb}
\usepackage[dvipdfm]{graphicx}
\usepackage[dvipdfm]{graphicx}
\usepackage{cancel}
\newcommand\sh{\operatorname{sinh}}
\newcommand\cth{\operatorname{coth}}
\newcommand\ptl{\partial}
\newcommand\tl{\tilde}
\newcommand\wtl{\widetilde}
\newcommand\vep{\varepsilon}
\newcommand\omH{\omega^{}_{_{\!\!H}}}
\newcommand\xms{|\chi|^2}
\newcommand\xrs{|\chi(\vec r\,)|^2}

\newcommand{\bb}{\begin{eqnarray}}
\newcommand{\ee}{\end{eqnarray}}

\begin{document}

%\Info
%Translated from Teoreticheskaya i Matematicheskaya Fizika,
%Vol.~164, No.~1, pp.~157--171, July, 2010.
%Original article submitted November 18, 2009.
%\endInfo

\begin{center}
{\large\bf THE ENERGY LEVEL SHIFTS, WAVE FUNCTIONS\\ AND THE
PROBABILITY CURRENT DISTRIBUTIONS\\ FOR THE BOUND SCALAR AND
SPINOR PARTICLES\\ MOVING IN A UNIFORM MAGNETIC FIELD}
\end{center}

\vskip3mm
\begin{center}{\bf V.~N.~Rodionov,}\footnote{Russian State
Geological Prospecting University, Moscow, Russia,\\ e-mail:
vnrodionov@mtu-net.ru.}\ {\bf G.~A.~Kravtsova,}\footnote{Lomonosov
Moscow State University, Moscow, Russia, e-mail: krav@dio.ru.}
\end{center}

\vskip5mm
\begin{center}
\parbox{110mm}{\sl\noindent
\hspace{10mm}We discuss the equations for the bound one-active
electron states based on the analytic solutions of the Schrodinger
and Pauli equations for  a uniform magnetic field and a single
attractive $\delta({\bf r})$-potential. It is vary important that
ground electron states in the magnetic field differ essentially
from the analogous state of spin-$0$ particles, whose binding
energy was intensively studied more than forty years ago. We show
that binding energy equations for spin-$1/2$ particles can be
obtained without using the language of boundary conditions in the
$\delta$-potential model developed in pioneering works. We use the
obtained equations to calculate the energy level displacements
analytically and demonstrate nonlinear dependencies on field
intensity. We show that the magnetic field indeed plays a
stabilizing role in considered systems in a case of the weak
intensity, but the opposite  occurs in the case of  strong
intensity. These properties may be important for real quantum
mechanical fermionic systems in two and three dimensions. We also
analyze the exact solution of the Pauli equation for an electron
moving in the potential field determined by the three-dimensional
$\delta$-well in the presence of a strong magnetic field. We
obtain asymptotic expressions for this solution for different
values of the problem parameters. In addition, we consider
electron probability currents and their dependence on the magnetic
field. We show that including the spin in the framework of the
nonrelativistic approach allows correctly taking the effect of the
magnetic field on the electric current into account. The obtained
dependencies of the current distribution, which is an
experimentally observable quantity, can be manifested directly in
scattering processes, for example.}
\end{center}


\section{  Formulation of the problem}

The effect of an external electromagnetic field on nonrelativistic
charged particles systems (such as atoms, ions, and atomic nuclei)
has long been investigated systematically (see, e.g.,~\cite
{Fr}-\cite {Man} ). Although this problem has a long history, a
set of questions still requires additional study. For example, a
systematic analysis of the bound states of spin-1/2 particles in
an intense magnetic field has been lacking until now. We note that
the basic results in the case of spinless particles were obtained
using analytic solutions in nonperturbative mathematical
treatments. As usual, the exact solutions of Schr\"{o}dinger
equations with Hamiltonians taking a particle bound by short-range
potential in the presence of external fields into account are
used. Furthermore, there is a rather common opinion that the role
of the magnetic field in decays of quasistationary states is
invariably stabilizing~\cite {Zel},~\cite{ Pop},~\cite{ Man}. This
view arises because the spinor states of electrons in an external
electromagnetic field are usually neglected in nonrelativistic
treatments, which is often inadequate~\cite {RoKr1},~\cite{RoKr2}.
In this paper, we treat an essential part of these problems.

We consider  charged spin-$0$ and spin-$1/2$ particles bounded by
a short-range potential ($\delta$ potential) and located in an
external stationary magnetic field with an arbitrary intensity. We
note that the zero-radius potential is a widespread approximation
for a multielectron atom field and especially for a negative ion
field~\cite {Baz},~\cite{Dem.3}. Energy level displacements can be
seen for the particle in a $\delta$ potential and a magnetic
field. The binding-energy equation is  most appropriate for the
investigating such states(\cite {Dem.1},~\cite{ Baz},~\cite{Lan}).

When an electron moves in a uniform magnetic field  oriented in
the  $z$ direction, the quantum mechanical system is invariant
with respect to the $z$ axis. The system  then becomes essentially
two-dimensional in the $xy$ plane. Many physical phenomena in
axially symmetric quantum systems of electrically charged fermions
(the quantum Hall effect~\cite {Pra}, high-temperature
superconductivity~\cite {Wil}, various film defects~\cite
{ChSolid}, etc.) can be effectively studied using the
nonrelativistic equations of motion in 2+1 dimensions. A number of
effects in constant magnetic fields, including  certain types of
doped two-dimensional semimetals, can be described using the Dirac
equation in 2+1 dimensions~\cite{Sc},~\cite{Ne}. But there are
many physical phenomena that occur in three-dimensional
space~\cite{Ter},~\cite{TeBag}. In this paper, we investigate the
effect of a stationary uniform magnetic field on localized
electron states in $2+1$ and 3+1 dimensions (see  also
~\cite{RodPhysRev},~\cite{TMP164}).

The effects of external electromagnetic fields on bound
nonrelativistic charged particles have already been studied
systematically in detail over several decades (see,
e.g.,~\cite{Man}, \cite{RoKr1}, \cite{Pop1}). But although this
problem has a very long history, several problems still need
additional study. In particular, such problems include analyzing
the effect of a strong magnetic field on bound charged particles
with their spin states taken correctly into account.

There is a widespread opinion that consistently including the
charged-parti\-cle spin is required only in studying relativistic
effects. But this is not always the case for a strong magnetic
field, as shown in~\cite{RoKr1},~\cite{RodDokl}. We note that
efforts to analyze spin effects in the nonrelativistic
approximation were made many times. Models precisely taking the
effect of the field on the particle into account were used for
this purpose. We emphasize that the wide experience in describing
spinless particles bound by the $\delta$-like potential in strong
electromagnetic fields has been acquired precisely because the
nonperturbative mathematical approach was used. Developed methods
based on the exact analytic solutions of the Schr\"odinger
equation~\cite{Man},~\cite{Pop1},~\cite{DemDr} were also used to
study processes taking particle spins into account. But because
mathematical estimates are complicated in the case of spinning
particles, the obtained conclusions are not always sufficiently
convincing. In our opinion, the
papers~\cite{RodPhysRev},~\cite{RodDokl},~\cite{RKMSakh}, where
the effects of the field on the behavior of spinning and scalar
particles were compared, are most correct from the standpoint of
the necessity to take particle spins into account. For these
purposes, the solutions of the Pauli equations, together with the
solutions of the Schr\"odinger equation, were also used in the
indicated papers. We here consider the effects produced by the
action of a strong constant uniform magnetic field on an electron
bound by a short-range potential.

 Under the action of the field, electron
energy levels are shifted by a quantity determined from the
transcendental equation for the energy. We note that this problem
was analyzed in the case of a scalar particle in~\cite{DemDr}. A
similar problem for the electron in the magnetic field with its
spin states taken into account was correctly solved
in~\cite{RodPhysRev} quite recently.

Our main purpose  is to derive equations for the binding energy of
a fermion in a field containing an attractive singular potential
and a stationary external magnetic field in the two- and
three-dimensional cases. We apply standard quantum mechanical
methods using the expansion of the unknown wave function  in a
series at the eigenfunctions obtained for the fermionic system in
the pure magnetic field. This method was used to study some
physical examples of the effect of a constant magnetic field on
charged particles bound by a single attractive $\delta$
potential~\cite{Kh1},~\cite{Kh2}. This formalism differs in
principle from the traditional derivation of wave functions in
similar problems using the boundary condition typical for the
$\delta$
potential~\cite{Dem.1},~\cite{Dem.2},~\cite{Pop},~\cite{Man}.

It is very important that our approach permits developing a
consistent investigation of the spin effects arising in an
external magnetic field.
  An exact analytic expression for
the wave function of a charged scalar particle in a state bound by
the $\delta$-potential and moving in a strong magnetic field was
found in~\cite{DemDr} using the Green's function of the
corresponding Schr\"odinger equation. We note that the Green's
function of a scalar particle in an external magnetic field was
presented in the classic monograph~~\cite{FH} (also
see~\cite{Pop}). As is known, the electron spin can be taken into
account, for example, by passing to the nonrelativistic limit in
the solutions of the Dirac equations describing the motion of a
spinning particle in a given external field~\cite{RMBull},
\cite{TMP145}.

The general structure of the paper and its main results are as
follows. In the next section, we construct the equation for scalar
particles with a low binding energy in a stationary external
magnetic field based on an  explicit solution of the
Schr\"{o}dinger equation. In Sec.~III, we obtain the expressions
for the energy of electron bound states in the $\delta$ potential
and an external magnetic field based on analogous analysis of
explicit solutions of the  Pauli equation. In Sec.~IV, we discuss
the equations for the binding energy of spin-$0$ and spin-$1/2$
particles in the presence of the both weak and strong magnetic
fields  because particle spin was previously taken into account
inadequately in similar problems.

Furthermore, in Sec.~V, we analyze the exact solution of the Pauli
equation for an electron moving in the potential well determined
by the three-dimensional $\delta$-function in the presence of a
strong magnetic field and to obtain asymptotic expressions for
this solution in the case of different values of the problem
parameters. In addition, in Sec.~VI we consider the probability
currents for the given particle and their dependence on the spin
and magnetic field.

\section{A scalar particle in an attractive potential in the presence
of a uniform magnetic field}

We consider a charge  in a uniform magnetic field ${\bf B}$
specified as \bb
 {\bf B}=(0,\,0,\,B)=\nabla\times {\bf A},\ \
 {\bf A}=(-yB,\,0,\,0).
\label{e1}
 \ee
 The Schr\"{o}dinger equation in field (\ref{e1}) has the form

\bb
 i\hbar\frac{\partial}{\partial t}\psi({t, \bf r})={\cal
H}\psi({t, \bf r}),\quad {\bf r}=(x, y,z), \label{eq12} \ee with
the Hamiltonian ${\cal H}$ is
 \bb {\cal H} =
\frac{1}{2m}\left(-i\hbar\frac{\partial}{\partial
x}+\frac{eB}{c}y\right)^2
-\frac{\hbar^2}{2m}\frac{\partial^2}{\partial y^2} -
\frac{\hbar^2}{2m}\frac{\partial^2}{\partial z^2}, \label{e12} \ee
where  $m$ and $e$ are the particle mass and charge. The particle
wave function in field (\ref{e1}) has the form~\cite{Lan}
 \bb
 \psi_{n p_x p_z}(t, {\bf r})=\frac12 e^{-iE_{n}t/\hbar}e^{i x p_x/\hbar + i z p_z/\hbar
 }U_n(Y),
\label{sol1} \ee where \bb E_{n}=\hbar \omega
\left(n+\frac12\right)+ \frac{{p_z}^2}{2m} \label{e2} \ee is the
electron energy spectrum, $\omega=|eB|/mc$, and $p_x$ and $p_z$
are the electron momenta  in the $x$ and $z$ directions.

The functions
$$
 U_n(Y) = \frac{1}{(2^n!\pi^{1/2}r_0)^{1/2}}
\exp\left(-\frac{(y-y_0)^2}{2r_0^2}\right)H_n\left(\frac{y-y_0}{r_0}\right),
$$
are expressed in terms of the Hermite polynomials $H_n(z)$, the
integer $n=0, 1, 2, \dots$ indicates the Landau level number,
$r_0=\sqrt{\hbar c/|eB|}\equiv \sqrt{\hbar/m\omega}$ is the
so-called magnetic length (see, for example,~\cite{Ter}) and
$y_0=-cp/eB$.

We now study a simple solvable model. We consider the motion of a
scalar particle in the three-dimensional case  in a single
attractive $\delta({\bf r})$-potential, where $\delta({\bf r})$ is
the Dirac delta function, in the presence of a uniform magnetic
field. In fact, we must solve the Schr\"odinger equation

\bb \frac{1}{2m}\left[\left(-i\hbar\frac{\partial}{\partial
x}+\frac{eB}{c}y\right)^2 -\hbar^2\frac{\partial^2}{\partial y^2}
-\hbar^2\frac{\partial^2}{\partial z^2} -\hbar^2\delta({\bf
r})\right]\Psi_{E'}({\bf r})=E'\Psi_{E'}({\bf r}). \label{e21} \ee
 We can take solutions of Eq.(\ref{e21})  in the form

\bb \Psi_{E'}({\bf r})= \sum\limits_{n,p_x,p_z} C_{E'np_x
p_z}\psi_{n p_x p_z}({\bf r})\equiv  \sum\limits_{n=0}^{\infty}
\int dp_x dp_z C_{E'np_x p_z}\psi_{np_x p_z}({\bf r}),
\label{sum1} \ee where $\psi_{n{p_x}{ p_z}}({\bf r})$ is the
spatial part of the wave functions (\ref{sol1}).

The coefficients $C_{E' n p_x p_z}$ can be easily calculated, and
we then obtain the equation

\bb 1= N \sum\limits_{n=0}^{\infty}\int dp_z\frac{1}{n+A},
\label{ener}
 \ee
 where $ N  $ is a  normalized coefficient
independent of the field and \bb
A=\frac12-\frac{E}{\hbar\omega}+\frac{{p_z}^2}{2 m
{\hbar\omega}}\label{ener1}. \ee Integrating over $p_z$ gives
(\ref{ener}) in the form
\begin{equation}
1=  N \pi \sqrt{2 m
\hbar\omega}\,\sum\limits_{n=0}^{\infty}\,\,\,\frac{1}{(n+A)^{1/2}}.
\label{cof}
\end{equation}

It is easy to see that Eq.(\ref{cof}) implicitly defines the
energy of a bound localized electron state in the magnetic field.
We note that (\ref{cof}) is consistent with analogous result in
\cite{Kh2}, where this equation was solved numerically. But Eq.
(\ref{cof}) can be analytically reduced to a simpler form. Indeed,
we can sum over $n$ in the right-hand side of Eq.(\ref{cof}) using
the representation \bb
\frac{1}{(n+A+i\varepsilon)^{1/2}}=\frac{e^{-i\frac{\pi}{4}}}{\sqrt{\pi}}\,\,\,
\int\limits_{0}^{\infty}\frac{e^{i(n+A+i\varepsilon)}}{t^{1/2}}dt.
\label{cof1} \ee As a result, Eq.(\ref{cof}) becomes \bb 1 = N_1
\sqrt{\hbar\omega}\,\,\,\frac{e^{-i\frac{3
\pi}{4}}}{2\sqrt{\pi}}\int\limits_{0}^{\infty}\frac{e^{-i\frac{E'
}{\hbar \omega}t}}{t^{1/2}\sin(t/2)}dt, \label{cof2} \ee where
$N_1$ is a real constant independent of the field. Because the
required energy
 \bb
 E'= - |E'| ,\label{dop1}
 \ee
must be negative, we can rotate the integration contour through
the angle $\pi/2$ in the complex $t$ plane. We thus obtain the
real expression \bb -1 = N_1
\frac{\sqrt{\hbar\omega}}{2\sqrt{\pi}}\int\limits_{0}^{\infty}\frac{e^{-\frac{E
}{\hbar \omega}t}}{t^{1/2}{\rm sinh}(t/2)}dt, \label{cof21} \ee
where  $E = |E'| \geq 0$. If we eliminate the magnetic field, then
(\ref{cof21}) takes the form

\bb -1 = N_1
\sqrt{\frac{\hbar}{\pi}}\int\limits_{0}^{\infty}\frac{e^{-{E_0 t
/\hbar}}}{t^{3/2}}dt, \label{cof3} \ee where $E_0=|E'_0|$ is the
absolute value of the binding energy of the particle in the
$\delta$-potential without the action of the external field.
Subtracting (\ref{cof3}) from (\ref{cof21}) and removing the
integral divergences in the lower limit by the standard
regularization procedure, we obtain \bb
      \int\limits_{0}^{\infty}\frac{e^{-E_0 t/\hbar}-e^{-E t/\hbar}}{t^{3/2}}dt= \int\limits_{0}^{\infty}\frac{e^{-E t/\hbar}}{t^{3/2}}\left( \frac{a_1 t}{{\rm {sinh}}{(a_1 t)}}-1\right)dt, \label{cof4}
\ee where $a_1=\frac{\omega}{2 } $. From (\ref{cof4}), it is easy
to obtain

\bb
\sqrt{E}-\sqrt{E_0}=\frac{\sqrt{E}}{2\sqrt{\pi}}\int\limits_{0}^{\infty}\frac{e^{-
x}}{x^{3/2}}\left( \frac{a x}{{\rm {sinh}}{(a x)}}-1\right)dx,
\label{cof5} \ee where $a=\frac{\hbar\omega}{2 E } $, which is
consistent with the analogous equation obtained by the well-known
method using  boundary conditions of wave functions in the
$\delta$-potential model \cite{Dem.2},\cite{Pop},\cite{RKMSakh}.

Expanding  the integrand function in (\ref{cof5}) in the
weak-field limit $\hbar\omega \ll 2 E_0$, we obtain
\bb
       E = E_0 \left(1-\frac{1}{48}\frac{{\hbar}^2{\omega}^2}{{E_0}^2} +
       \frac{1}{576}\frac{{\hbar}^4{\omega}^4}{{E_0}^4}\right). \label{cof51}
\ee We note that the quadratic term  in (\ref{cof51}) coincides
with analogous result in \cite{Dem.2}.

To consider the strong-field case $\hbar\omega > 2 E_0$, we reduce
the right-hand side of (\ref{cof5}) to the analytic form
\bb
        -\frac{1}{\sqrt{a_0}}=\frac{1}{\sqrt{2}}\zeta\left(\frac{1}{2},\frac{1}{2}
        +\frac{1}{2 a}\right), \label{cof6}
\ee where $ a_0 = \frac{\hbar\omega}{2 E_0}$ and $\zeta[\nu,p]$ is
the Hurwitz (generalized Riemann) zeta function. The validity
range of (\ref{cof6}) can be found somewhat wider that was
initially assumed. In deriving (\ref{cof21}) we assume that $E'
\leq 0$, but we can see from (\ref{cof6}) that argument of zeta
function can continuously reach the values $$ 1/2 + 1/2a > 0 .
$$
This condition limits the required binding-energy spectrum by \bb
   E' < \frac{\hbar\omega}{2}. \label{cof61}
\ee The physical meaning of this condition is the  restriction to
the continuous spectrum of the scalar particle in the magnetic
field by the value of (\ref{cof61}) (i.e., $ E' \ge \hbar\omega
/2$).
  We note that
after a change of variables in (\ref{cof6}), it is coincides with
the basic equation in \cite{Dem.2}, where the case of scalar
particles in the magnetic field was considered and an analogous
conclusion about limitation of continuous spectrum was drawn.

Expanding  $\zeta(\nu,p)$ for $ p \ll 1$ gives
 \bb
   \zeta(1/2,p)=\frac{1}{p^{1/2}}+\zeta(1/2)-\frac{1}{2}\zeta(3/2)p
   +\frac{3}{8}\zeta(5/2)p^2 + 0[p]^3. \label{cof7}
\ee
Substituting (\ref{cof7}) in (\ref{cof6}), we  obtain the
explicit equation for the bound-state energy in the strong-field
limit

\bb
   {E'}=  \hbar\omega\left( 0.205-0.452\sqrt{\frac{E_0}{\hbar\omega}}
   -0.367\, \frac{E_0}{\hbar\omega}\right).\label{cof8}
\ee
We emphasize that in superstrong  magnetic fields, expansion
(\ref{cof8}) gives the upper limit
$$
           E' =  0.205\, \hbar\omega\,,
           $$
for the binding energy of the scalar particle, which does not
contradict condition (\ref{cof61}). Furthermore, it can be seen
that this limiting value is independent of the particle energy in
the absence of the field and is completely determined by the
magnetic field intensity.

It is interesting to compare the obtained results with the results
in the case of two-dimensional model. The analogue of
Eq.(\ref{cof}) in the two-dimensional case is

\bb
   1=  \frac{1}{8\pi}\int\limits_{0}^{\infty}\frac{e^{-E t/\hbar\omega}}{{\rm sinh}(t/2)}dt, \label{cof9}
\ee
 which coincides with the corresponding result in
\cite{Kh1}. But our regularization procedure  here essentially
differs from \cite{Kh1}. As before (see (\ref{cof4})), we remove
the magnetic field and obtain \bb 1=
\frac{1}{4\pi}\int\limits_{0}^{\infty}\frac{e^{-E_0 t}}{t}dt.
\label{cof10} \ee With a simple calculation similar to that
 in  the three-dimensional case, we can write
 \bb
     \ln\frac{E}{E_0}=\int\limits_{0}^{\infty}\frac{e^{-x}}{x}\left( \frac{a x}{{\rm sinh}(a x)}-1 \right) dx, \label{cf1}
\ee where  $a=\hbar\omega/(2 E)$ as before. In the weak-field
limit, we obtain \bb
      E=E_0 \left(1- \frac{\hbar^2\omega^2}{24 {E_0}^2}\right)\label{cf2}
\ee from (\ref{cf1}).

To consider the range $\hbar\omega > 2 E_0$, we must first
calculate  the integral in the right-hand side of  Eq.(\ref{cf1})
analytically,

\bb -\ln\left( \frac{E}{E_0} \right)= \ln(2 a)+
\Psi\left(\frac{1+a}{2a} \right), \ee
 where $\Psi(x)$ is a logarithmic derivative of Euler gamma function.
 We then have the basic equation in the two-dimensional model
 \bb
- \ln\left( 2 a_0 \right)=\Psi\left( \frac{1}{2}+\frac{1}{2 a}
\right), \label{cf3}
 \ee
where $a_0=\hbar\omega/(2 E_0).$

 In the strong-field limit, after evaluation the function
 $\Psi(p)$,
\bb \Psi (p) =-\frac{1}{p} -C +\frac{\pi^2}{6} p + \frac{1}{2}{
\Psi}^{(2)}(1) p^2 +\frac{\pi^4}{90} p^3 +0[p]^4, \label{dop2} \ee
where
$$
{ \Psi}^{(n)}(z) = \frac{d^n }{d z^n}\Psi(z) = {(-1)}^{n+1} n!
\zeta (n+1, z),
$$
we can write  (\ref{cf3}) as
 \bb
 \ln{\frac{\hbar\omega}{E_0}} - \frac{1}{\frac{1}{2}-\frac{E'}{\hbar\omega}}- C +\frac{\pi^2}{6}
 \left( \frac{1}{2}-\frac{E'}{\hbar\omega}\right)= 0, \label{cf41}
 \ee
 where  $C = 0.577...$ is the Euler constant.

The solution of Eq.(\ref{cf41}), which explicitly determines the
bound-state energy, can be written as \bb
\frac{E'}{\hbar\omega}=\frac{1}{2}- \frac{6 (C
-\ln(\hbar\omega/E_0))+\sqrt{24 \pi^2 +36(C
-\ln(\hbar\omega/E_0))^2}}{2 \pi^2}.\label{cf5} \ee For
$\ln(\hbar\omega/E_0)\gg 1 $  we obtain  \bb
\frac{E'}{\hbar\omega}
=\frac{1}{2}-\frac{1}{\ln(\hbar\omega/E_0)}- \frac{C
}{\ln^2(\hbar\omega/E_0)}+\frac{\left( \frac{\pi^2}{6}-C
\right)}{\ln^3(\hbar\omega/E_0)} + 0[\ln(\hbar\omega/E_0)]^4
\label{dop3} \ee from (\ref{cf5}). Considering the properties of
$\Psi(z)$, we see that expansion (\ref{dop3}) is correct for the
binding energy $ E' =  \le \hbar\omega /2$. Furthermore, this
limiting value, as before, is independent of the particle energy
in the absence of the field.  But there is an essential difference
from the three-dimensional case. For supperstrong magnetic fields
(when not only $\hbar\omega/E_0$ is large but also
$\ln(\hbar\omega/E_0)
>>1)$, the upper
limit of the shifted binding-energy level in the considered model
tends directly to the boundary of the continuous spectrum.

\section{A spin particle in an attractive potential in the presence
of a uniform magnetic field}

It is vary important that we can use the present approach to study
the spin effects in magnetic fields in the same way. The case of a
spin-$1/2$ particle can be calculated based on exact solutions of
the Pauli equation. The Pauli equation in the field (\ref{e1}) has
the form

\bb i\hbar\frac{\partial}{\partial t}\psi({t, \bf r})={\cal
H}\psi({t, \bf r}),\quad {\bf r}=(x, y,z), \label{eq12} \ee with
the Hamiltonian ${\cal H}$ is \bb {\cal H} =
\frac{1}{2m}\left(-i\hbar\frac{\partial}{\partial
x}+\frac{eB}{c}y\right)^2
-\frac{\hbar^2}{2m}\frac{\partial^2}{\partial y^2} -
\frac{\hbar^2}{2m}\frac{\partial^2}{\partial z^2}+ \mu\sigma_3B,
\label{e12} \ee where  $\mu=|e|\hbar/2m c$ is the Bohr magneton,
$m$ is the mass of a  electron and
$$
\sigma_3=\left(\begin{array}{cc}
1 & 0\\
0 &-1\\
\end{array}\right)
$$
is the  $z$-component of Pauli matrices. The last term in
(\ref{e12}) describes the interaction of the electron spin
magnetic moment with the magnetic field. The electron wave
function in field (\ref{e1}) has the form \bb
 \psi_{n p_x p_z s}(t, {\bf r})= \frac{1}{2}\psi_{n p_x p_z}(t, {\bf r})
\left( \begin{array}{c}
1+s\\
1-s
\end{array}\right),
\label{sol2} \ee
 where $\psi_{n p_x p_z}(t, {\bf r})$ is the solution of the Schr\"{o}dinger equation in
 field (\ref{e1}) (see(\ref{sol1})),
 \bb E_{ns}=\hbar \omega
\left(n+\frac12\right)+ \frac{{p_z}^2}{2m}+ s\hbar
\omega\frac{1}{2} \label{e2} \ee is the electron energy spectrum,
$\omega=|eB|/mc$, $s=\pm 1$ is the conserved spin quantum number,
and  $p_x$ and $p_z$ are the electron momenta in the $x$ and $z$
directions.

It is very important that the electron ground state in a  magnetic
field differs essentially from the analogous state of spin-$0$
particles. Moreover, the continuous spectrum boundaries differ for
spinor and  scalar particles. For example, if the continuous
spectrum of a scalar particle begins from $E' \geq \hbar
\omega/2$, then the continuous spectrum of an electron begins from
$E'\geq \hbar \omega$ for the spin directed along the magnetic
field direction and from $E'\geq 0$ for the spin directed against
the magnetic field direction.

Taking the interaction of the electron spin magnetic moment with
the magnetic field into account, we can write the energy equation
in the three-dimensional case in the form

 \bb
 \sqrt{E}-\sqrt{E_0}= \frac{\sqrt{E}}{2\sqrt{\pi}}\int\limits_{0}^{\infty}\frac{e^{-t }}
 {t^{3/2}}\left(\frac{a t e^{- s a t}}{{\rm sinh}(a t)}-1\right)dt, \label{cf6}
\ee were $s=\pm 1$ respectively corresponds to spin orientations
 along or against
the magnetic field direction. Expanding the integral in
(\ref{cf6}) for $a << 1$, we obtain the equation
\bb \sqrt{E}
-\sqrt{E_0}=
-\frac{s\hbar\omega}{4\sqrt{E}}+\frac{\sqrt{E}}{12}\left(\frac{\hbar\omega}{2
E_0}\right)^2.\label{cf62}
\ee
The solution of Eq.(\ref{cf62}) has
the form
\bb
           \sqrt{\frac{E}{E_0}}=\frac{2\left( -12 E_0 -\sqrt{3}\sqrt{48 {E_0}^4+
           s(\hbar^3\omega^3 E_0 -48 \hbar\omega {E_0}^3}\right)}{\hbar^2\omega^2 -48 {E_0}^2}.\label{cf63}
\ee Expansion  (\ref{cf63}) can be written in the weak-field limit
as

\bb
 \frac{E}{E_0}=1 - s\frac{\hbar\omega}{2 E_0}-\frac{1}{48} \frac{\hbar^2\omega^2}{E_0^2}. \label{cf61}
\ee

It is easily seen from (\ref{cf61}) that the energy level
$E'_0=-|E_0|$ existing in the $\delta$-potential without a
perturbation is shifted under the magnetic field action by
$+\frac{\hbar\omega}{2 E_0}$ (upward) in the case $s=1$ and by
$-\frac{\hbar\omega}{2 E_0}$ (downward) in the case $s=-1$. But
the depth of the arrangement of energy levels with respect to the
continuous spectrum boundaries  is the same in these two cases. We
note that it has the same depth in the case of spin-0 particles.

Integrating of the right-hand side Eq.(\ref{cf6}), we obtain the
equation in the analytical form \bb
   -\frac{1}{\sqrt{a_0}}=\frac{1}{\sqrt{2}} \zeta \left[\frac{1}{2},\frac{1}{2}+
   \frac{s}{2}+\frac{1}{2 a}\right]. \label{cf7}
\ee In the strong-field limit $ \hbar\omega > E_0$,  we can write
the Hurwitz zeta-function in Eq.(\ref{cf7}) as \bb
      \frac{1}{\sqrt{2}} \zeta \left[\frac{1}{2},\frac{1}{2}+\frac{s}{2}+\frac{1}{2 a}\right]= \frac{1}{\sqrt{2}}\frac{1}{\sqrt{\frac{1+s}{2}+\frac{E}{\hbar\omega}}}
      +\frac{\zeta[1/2]}{\sqrt{2}}-\frac{\zeta[3/2]}{2\sqrt{2}}\left(\frac{1+s}{2}
      +\frac{E}{\hbar\omega}\right).
\ee Finally, we write the Eq.(\ref{cf7})as
\bb
    \frac{1}{\sqrt{2}} +\left( \sqrt{\frac{2 E_0}{\hbar\omega}}+\frac{\zeta [1/2]}{\sqrt{2}}\right)x -
    \frac{\zeta [3/2]}{2 \sqrt{2}} x^3 = 0, \label{cf8}
\ee where $$x = \sqrt{\frac{1+s}{2}+\frac{E}{\hbar\omega}}.$$ The
solutions of Eq.(\ref{cf8}) for different spin values $ s =0, \,
+1 ,\, -1$ can be  represented as

  \bb
      E' =  \hbar \omega \left( 0.205 + \frac{s}{2} -0.452 \sqrt{\frac{E_0}{\hbar\omega}}-
      0.367 \frac{E_0}{\hbar\omega}  \right). \label{cf81}
\ee In  Fig.~1, we show graphs of solutions of  Eq.(\ref{cf7}) for
different particle spin values with $E_0 = 1$ and $\hbar\omega =
100 E_0$, which are typical values for experiments on the
bound-state energy in semiconductors at low temperatures in
magnetic fields $H \sim 10^5 {\rm G}$.
\begin{figure}[h]
\vspace{-0.2cm} \centering
\includegraphics[angle=270, scale=0.5]{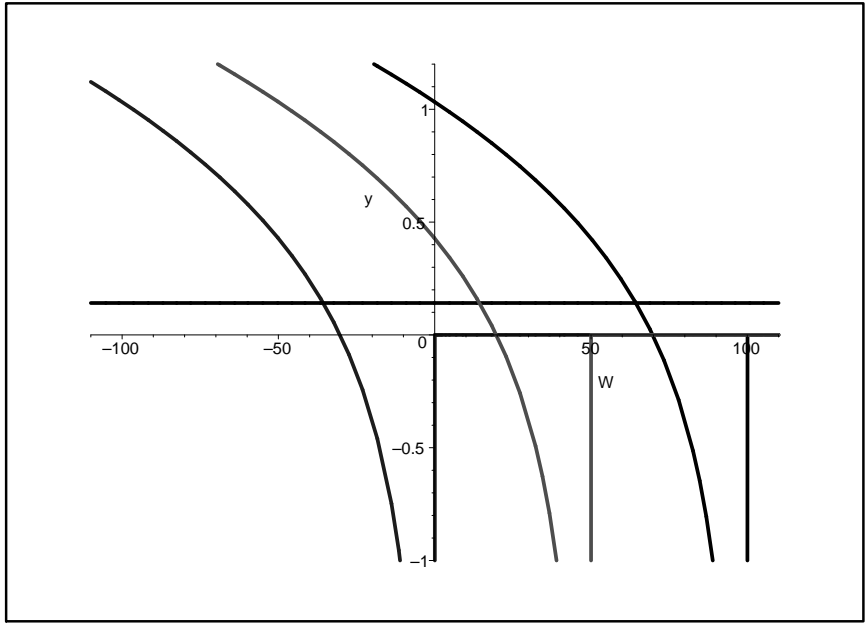}
\caption{Solutions of Eq.(\ref{cf7})for different particle spin
values. The horizontal straight line is $y ={a_0}^{-1/2}$. The
curves are $y=-2^{-1/2}\zeta[1/2,(1+s)/2+1/2a]$ and from left to
right $s=-1,0,1$. The parameter $W=E'/E_0$ is also shown.}
\vspace{-0.1cm}
\end{figure}

It is obvious  that the approximate solutions of (\ref{cf81}) are
quite near the intersections of the graphs of left-hand side of
Eq.(\ref{cf7}) (the straight line in Fig.~1) and of the right-hand
sides of Eq.(\ref{cf7}) (the curves in Fig.~1). We emphasize that
the dependence of energy level shifts on the particle spin does
not disappear in the strong-field limit. Moreover, the continuous
spectrum boundaries are shifted in the cases $s=0$ and $s=1$.
Hence, the perturbative displacements of the binding-energy levels
(as in the weak-field limit) are at the same distances from the
continuous spectrum boundaries  in all cases.

We now consider spin interactions in the two-dimensional case.
According to our approach, we can write

\bb \ln
\left(\frac{E}{E_0}\right)=\int\limits_{0}^{\infty}\frac{e^{-x}}{x}
\left( \frac{a x e^{- s a x}}{{\rm sinh}(a x)} -1\right) dx,
\label{cf91} \ee where the particle spin  direction, as before, is
represented by $s=\pm 1$. In the weak-field limit, we obtain
 \bb
                  \frac{E}{E_0}= 1-\frac{s}{2}\frac{\hbar \omega}{E_0}-
                  \frac{1}{24}\left( \frac{\hbar \omega}{E_0}\right)^2 \label{cf9}
\ee from Eq.(\ref{cf91}). To consider the strong-field limit, we
must
 calculate the integral in Eq.(\ref{cf91}) for $s=\pm 1$ in analytic form
 \bb
\int\limits_{0}^{\infty}\frac{e^{-x}}{x} \left( \frac{a x e^{- s a
x}}{{\rm sinh}(a x)} -1\right) dx = -2a\frac{(1+
s)}{2}-\ln(2a)-\Psi\left( \frac{1}{2 a }\right). \label{cf10}
 \ee
 We can then write the equations for energy displacements for
$\hbar\omega > E_0 $ as \bb
          \ln\left( \frac{E}{E_0} \right)= 2a\frac{(1-s)}{2}-\ln(2a)+ C -\frac{\pi^2}{12 a}. \label{cf11}
\ee
 For the case $s= -1$ in the strong-field limit ($ \ln\frac{\hbar\omega}{E_0} >> 1 $)
we immediately obtain
 \bb
    E' = -\hbar\omega\left(\frac{1}{\ln\frac{\hbar\omega}{E_0}}+\frac{C}{\ln\frac{\hbar\omega}{E_0}}\right)\label{cf101}
\ee from Eq.(\ref{cf91}).

For the opposite spin orientation ($ s=1$) in Eq.(\ref{cf10}), we
must first use the recurrent relation for $\Psi (p)$,

\bb
        \frac{1}{x} +  \Psi (x) = \Psi (1+ x).
\ee Using the asymptotic expansion for $ \Psi (p)$ (see
(\ref{dop2})), then obtain
\bb
        E' = \hbar\omega\left(1 - \frac{1}{\ln\frac{\hbar\omega}{E_0}}-\frac{C}{(\ln\frac{\hbar\omega}{E_0})^2}\right).\label{cf101}
\ee

The  dependence on the spin parameters can be interpreted  as in
three-dimensional model. We can write this equation in the form
\bb \ln\left( \frac{\hbar \omega}{E_0} \right)= - \psi \left(
\frac {1+s}{2} + \frac {1}{2a} \right) . \label{new1}\ee

In Fig.2, we show graphs of the solutions of Eq.(\ref{new1}) for
different particle spin values with $ E_0 = 10^{-3} eV$ and $\hbar
\omega = 100 E_0$. The main difference from the three-dimensional
model is that the perturbative binding-energy levels converge to
the continuous spectrum boundaries in a superstrong magnetic field
in this case.

\begin{figure}[h]
\vspace{-0.2cm} \centering
\includegraphics[angle=270, scale=0.5]{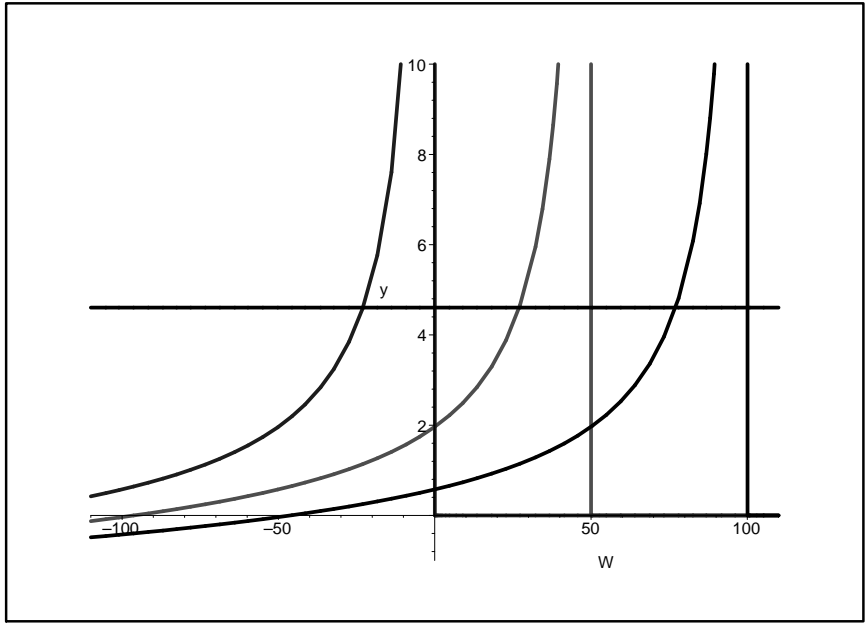}
\caption{Solutions of Eq.(\ref{new1})for different particle spin
values. The horizontal straight line is $y =\ln{\hbar\omega/E_0}$.
The curves are $y=-\Psi[(1+s)/2+1/2a]$ and from left to right
$s=-1,0,1$. The parameter $W=E'/E_0$ also shown.} \vspace{-0.1cm}
\end{figure}

\section{{The binding energy of
spin-$0$ and spin-$1/2$ particles in the presence of the both weak
and strong magnetic fields }}
%\label{s4}

We have shown that the effect of a magnetic field on localized
electron states leads to equations for the binding energy of
spin-$0$ and spin-$1/2$ particles. In the weak-field limit ($\hbar
\omega \ll E_0$), the energy displacements of scalar and spinor
particles are described by the expressions
 \bb
  \left. \begin{array}{rcl}
 s=0\,\,\,\,\,\, \frac{\hbar\omega}{2 E_0}&+&\frac{E}{E_0}\\
 s=1\,\,\,\,\,\,\,\,\frac{\hbar\omega}{E_0} &+& \frac{E}{E_0}\\
 s=-1\,\,\,\,\,\,\,\,\,\,\,\,\,\, &{}& \frac{E}{E_0}\\
  \end{array}\right \} = 1 + \frac{\hbar\omega}{2 E_0} - \frac{\hbar^2\omega^2}{48 {E_0}^2}. \label{car}
\ee in the three-dimensional case and
\bb
  \left. \begin{array}{rcl}
 s=0\,\,\,\,\,\, \frac{\hbar\omega}{2 E_0}&+&\frac{E}{E_0}\\
 s=1\,\,\,\,\,\,\,\,\frac{\hbar\omega}{E_0} &+& \frac{E}{E_0}\\
 s=-1\,\,\,\,\,\,\,\,\,\,\,\,\,\, &{}& \frac{E}{E_0}\\
  \end{array}\right \} = 1 + \frac{\hbar\omega}{2 E_0} - \frac{\hbar^2\omega^2}{24 {E_0}^2} \label{car1}
\ee in the two-dimensional case. From Eq.(\ref{car}) and
Eq.(\ref{car1}) we can also obtain

$$
s=0:\,\,\,\,\,\,\,\,\,\,\,\,\,\,\, E = E_0 -
\frac{\hbar^2\omega^2}{24 \delta {E_0}}, \label{}
$$

$$
 s=1:\,\,\,\,\,\,\,\,\, E = E_0 -\frac{\hbar\omega}{2} -
\frac{\hbar^2\omega^2}{24 \delta {E_0}}, \label{}
$$

 \bb
 s=-1:\,\,\,\,\,\,\,E = E_0 + \frac{\hbar\omega}{2}
- \frac{\hbar^2\omega^2}{24 \delta {E_0}},
 \label{car11}
 \ee
where  $\delta=2$ and $\delta=1$ in the three-dimensional and in
the two-dimensional cases correspondingly.

The dependence on the particle spin does not disappear in the
strong-field limit ($\hbar \omega \gg E_0$). In three-dimensional
 case, the perturbative energy levels approach specific spectral
 values determined by the magnetic field intensity. For different particle
spin values, the displacements of the binding-energy levels as
before [see Eq.(\ref{car})and Eq.(\ref{car1})] are at identical
distances from the continuous spectrum boundaries and can be
represented in the form
\bb
  \left. \begin{array}{rcl}
s=0\,\,\,\,\,\,\,\,\, \frac{\hbar\omega}{2}&-&E'\\
s=1\,\,\,\,\,\,\,\,\, \hbar\omega    & - &E'     \\
s=-1\,\,\, \,\,\,\,\,\,\,\,\,\,\,\,\,\,       &{-} & E'    \\
  \end{array}\right \} =  \hbar\omega\left( 0.295 +  0.452 \sqrt{\frac{E_0}{\hbar\omega}}+
  0.367 \frac{E_0}{\hbar\omega}\right). \label{car2}
\ee

Removing the braces in Eq.(\ref{car2}), we find

$$
s=0:\,\,\,\,\,\,\,\, E' =0.205 \hbar\omega - 0.452
\sqrt{E_0\hbar\omega}-
  0.367 E_0 , \label{}
$$

$$
 s=1:\,\,\,\,\,\,\,\,\, E' = 0.705 \hbar\omega - 0.452
\sqrt{E_0\hbar\omega}-
  0.367 E_0 , \label{}
$$

\bb
 s=-1:\,\,\,\,\,\,\,\,\,E' = -0.295 \hbar\omega - 0.452
\sqrt{E_0\hbar\omega}-
  0.367 E_0 .
\label{car12} \ee

In particular, from Eq.(\ref{car12}) one can immediately see that
the value of the binding energy level is positive both in the case
of a spinless particles ($s=0$) and in the case where the electron
spin in oriented along the magnetic field direction ($s=1$), but
it remains  negative in the case where the electron spin is
oriented against the magnetic field direction ($s=-1$).

It can be easily seen that the dependence on the spin parameters
in the two-dimensional case can be written analogously,
 \bb
  \left. \begin{array}{rcl}
s=0\,\,\,\,\,\,\,\,\, \frac{\hbar\omega}{2}&-&E'\\

s=1\,\,\,\,\,\,\,\,\, \hbar\omega    & - &E'     \\

s=-1\,\,\, \,\,\,\,\,\,\,\,\,\,\,\,\,\,       &{-} & E'    \\
  \end{array}\right \} =  \frac{\hbar\omega}{\ln\frac{\hbar\omega}{E_0}}+
  \frac{C\,\, \hbar\omega}{(\ln\frac{\hbar\omega}{E_0})^2} , \label{car3}
\ee and without braces we have

 $$
s=0:\,\,\,\,\,\,\,\, E' = \frac{\hbar\omega}{2} -
\frac{\hbar\omega}{\ln \left(\frac{\hbar\omega}{E_0}\right) } -
\frac{C\,\,
\hbar\omega}{{\ln}^2\left(\frac{\hbar\omega}{E_0}\right)}
,\label{} $$

$$ s=1:\,\,\,\,\,\,\,\,\, E' = \hbar\omega -
\frac{\hbar\omega}{\ln\left( \frac{\hbar\omega}{E_0}\right)} -
\frac{C\,\,
\hbar\omega}{{\ln}^2\left(\frac{\hbar\omega}{E_0}\right)}
,\label{} $$

\bb s=-1:\,\,\,\,\,\,\,\,\,E' = - \frac{\hbar\omega}{\ln\left(
\frac{\hbar\omega}{E_0}\right)} - \frac{C\,\,
\hbar\omega}{{\ln}^2\left(\frac{\hbar\omega}{E_0}\right)}.
\label{car13} \ee

It can be seen from Eq.(\ref{car3}) and Eq.(\ref{car13}) that the
energy levels in the basic terms are also independent on the
particle energy in the absence of the magnetic field. The
distinctive feature of this case is that the binding-energy levels
for superstrong magnetic fields [when $\ln(\hbar\omega/E_0)
>>1$] directly approach the continuous spectrum
boundaries for all considered spin values.

We have shown  that the energy levels of a polarized electron
under the action of a weak magnetic field for different particle
spin values are shifted similarly in the three-dimensional and
two-dimensional models. We have the line displacements as the
levels themselves for $s=1$ and $s=-1$ and analogous shifts of the
continuous spectrum boundaries of for $s=1$. We also have the same
picture in the case of a spinless particle with the line shift of
the continuous spectrum boundary. Clearly, in case of weak
intensity, a magnetic field indeed plays a stabilizing role in the
considered systems because the depth of the perturbative
binding-energy levels from the continuous spectrum boundaries are
shifted downward under the field action independently of the
particle spin. But our results show  a nonlinear dependence on the
field intensity in the strong-field limit. Nevertheless, the
continuous spectrum boundaries in the cases $s=0$ and $s=1$, as
before, have a linear dependence on the field in this limit. In
superstrong magnetic fields, the binding-energy levels can
approach the continuous spectrum boundaries. The distinctions can
be formulated as follows. In the three-dimensional model, there is
a fixed depth of the energy levels from the continuous spectrum
boundaries that is the same for all spin values. In the
two-dimensional model, the  energy levels in a superstrong
magnetic field tend asymptotically to the continuous spectrum
boundaries. But in both cases, the system instability increases
 in strong magnetic fields.
This conclusion therefore disproves the opinion that a magnetic
field always plays a stabilizing role in systems of bound
particles.

\section{{The exact analytic solution of the Pauli equation
for the attractive three-dimensional $\delta$-well and its
asymptotic expressions}}
%\label{s2}

 The  Green's function obtained in~\cite{RMBull}  is the
solution of the Pauli equation with a $\delta$-source and can be
represented as an integral over time~\cite{TMP145}, \cite{TMP164}.
This integral determining the Green's function admits a Wick
rotation to the lower complex half-plane of the variable $t$ (see,
e.g.,~\cite{Ter}). This operation makes the integral purely real.
As a result, the stationary Green's function can be represented in
the form
\begin{align}
G_\text{W}(\vec{r},\vec{0})={}&-i\biggl(\frac m{2\pi}\biggr)^{3/2}
\frac{\omH}2e^{-im\omH xy/4}\times{}
\nonumber
\\
&{}\times\int_0^{\infty}\frac{dt}{t^{1/2}}
\sh^{-1}\biggl(\frac{\omH t}2\biggr)e^{S/\hbar}
\begin{pmatrix}1/2+s\\[1mm]1/2-s\end{pmatrix},
%\label{1}
\end{align}
where $m$ is the particle mass, $\omH=eH/mc$ is the cyclotron
frequency, $e$ is the absolute value of the particle charge, $H$ is
the strength of the uniform magnetic field oriented along the $z$
axis, $s =\pm1/2$ is the spin number, and the argument of the
exponential in the integrand is in fact the classical action
function
\begin{equation}
S=-\frac{mz^2}{2t}-\frac14m\omH\rho^2
\cth\biggl(\frac{\omH t}2\biggr)+(W-s\cdot\hbar\omH)t
%\label{2}
\end{equation}
(although the argument of the exponential formally contains the
Planck constant, this dependence vanishes because of the shift in
the energy of a bound spinning particle in a magnetic
field~\cite{RodPhysRev}). In formulas~(1) and~(2), $x$, $y$, and
$z$ are the Cartesian coordinates, $\rho^2=x^2+y^2$, and $W<0$ is
the total energy of the bound particle. We note that~(1),~(2) is
the ordinary propagator of a charged particle moving in a magnetic
field and is continued to the range of negative energies.

We write the spatial part of the wave function in the form
\begin{equation}
\psi(\vec{r}\,)=\wtl{N}e^{-im\omH xy/4}
\int_0^{\infty}\frac{dt}{t^{1/2}}
\sh^{-1}\biggl(\frac{\omH t}{2}\biggr)e^{S/\hbar},
%\label{3}
\end{equation}
where $\wtl{N}$ is the normalizing coefficient. We pass to more
natural variables using characteristic scales of the problem. As
an energy scale, we take $|W_0|$, which is the absolute value of
the particle binding energy in the absence of a magnetic field. We
let $w=W/|W_0|$ denote the dimensionless binding energy in such
units. It is equal to $-1$ in the absence of the external field.
Of course, $w$ depends on both the external field $H$ and the spin
direction in the general case~\cite {RodPhysRev}. It is convenient
to measure the magnetic field in the dimensionless units
$h=\hbar\omH/ |W_0|$. The de~Broglie wavelength
$l_0=\hbar(2m|W_0|)^{- 1/2}$ of the particle for the zero magnetic
field can serve as a natural coordinate scale, and the quantity
$t_0=\hbar|W_0|^{- 1}$ can serve as a time scale. Using such units
for the spatial part of the wave function, we obtain
\begin{equation}
\psi(\vec{r}\,)=Ne^{-ih\tl x\tl y/4}\int_0^{\infty}\frac{d\tau}{\tau^{1/2}}
\frac{\exp\big[-{\tl z}^2/(4\tau)-h{\tl\rho}^2\cth(h\tau/2)/8-
\wtl w\tau\big]}{1-e^{-h\tau}},
%\label{4}
\end{equation}
where $N$ is the normalizing coefficient in the new variables and
$\tl x$, $\tl y$, $\tl z$, and $\tau$ are the dimensionless
coordinates and time. The analogy between the integrand in~(4) and
the classical Planck formula for the black-body emission spectrum is
interesting. The quantizing character of the magnetic field (as the
quantum character of radiation) is not manifested until the exponent
factor $h\tau$ in the denominator in the right-hand side of~(4) is
sufficiently small. We also call attention to the phase factor
$e^{-ih\tl x\tl y/4}$, whose presence explicitly demonstrates the
existence of the orbital probability current.

To calculate the normalizing coefficient in~(4), we must evaluate several
integrals. The integrals over the coordinates are assumed to be purely
Gaussian, and the integral over time can be reduced to the generalized
Riemann zeta function $\zeta(3/2,\vep)$ by a simple change of variables. As
a result, we obtain
\begin{equation}
N=\frac{h^{5/4}}{2\pi l_0^{3/2}}
\zeta^{-1/2}\biggl(\frac32,\frac{\wtl w}h\biggr),
%\label{5}
\end{equation}
where $\wtl w=w_{\mathrm{sh}}-w$ is the absolute value of the
so-called effective energy of the particle. The parameter $\wtl w$
appears because the lower edge of the continuum, together with the
bound states $w$, is shifted in the magnetic field. Consequently,
the binding {\sl effective energies} are now measured from new
boundaries determined by the expression
$w_{\mathrm{sh}}=(s+1/2)h$~\cite {RodPhysRev}. It also follows
directly from this relation that the shift $w_{\mathrm{sh}}$ in
the continuum depends explicitly on both the magnetic field and
the particle spin orientation. But it was shown in~\cite
{RodPhysRev} using weak ($h\ll1$) and strong ($h\gg1$) fields as
examples that $\wtl w$ is independent of the spin.

It is easy to see that the integral in~(4) can be related to
Laplace-type integrals~\cite {Fed},~\cite {Migd}. The relevant
integration domain is determined by a neighborhood of a single
point of the exponential maximum. We consider the contribution of
the saddle point to the integral in~(4) written in the form
\begin{equation}
\int_0^{\infty}\frac{du}{u^{1/2}}\frac{e^{-hg(u)}}{1-e^{-u}},\qquad
g(u)=\frac{{\tl z}^2}{4u}+\frac{\tl\rho^2}8
\cth\biggl(\frac{u}{2}\biggr)+\frac{\wtl w}{h^2}u.
%\label{6}
\end{equation}
Here, $u=h\tau$, and the saddle point is a root of the equation
\begin{equation}
Bu^2-A=\frac{u^2}{\sinh^2(u/2)},\qquad
A=4{\tl z}^2\tl\rho^{-2},\qquad B=16\wtl w\tl\rho^{-2}h^{-2}.
%\label{7}
\end{equation}

\begin{figure}[h]
\centering
\includegraphics{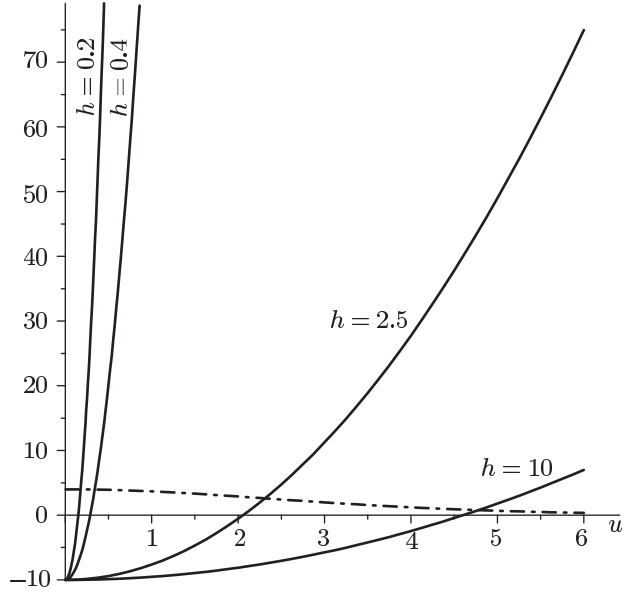}
\caption{The solution of transcendental equation~(7) in the plane $z=1.6$
for $\rho=1$. The solid curves correspond to the left-hand side of Eq.~(7)
for different values of the magnetic field, and the dot-dashed curve is the
graph of the right-hand side of Eq.~(7).}
%\label{fig1}
\end{figure}

A graphic illustration of the search for the solution of this
transcendental equation is shown in Fig.~3, which shows the family
of parabolas corresponding to the left-hand side of Eq.~(7) for
different values of the parameter $h$ (the solid curves) and the
graph of the function in the right-hand side of the equation (the
dot-dashed curve). It is easy to verify that the parabola branch
can intersect the monotonic function $u^2\sinh^{-2}(u/2)$ in
various ranges of the integration parameter $u$. Hence, in the
range $u\ll1$, the right-hand side of the equation differs
slightly from the constant, and the saddle point, which is the
root of the quadratic equation in this case, is given by
\begin{equation}
u_0=\frac{h\sqrt{\tl\rho^2+\tl z^2}}{2\sqrt{\wtl w}}.
%\label{8}
\end{equation}
In the other limit case $u\gg1$, the right-hand side of Eq.~(7) almost
vanishes, and for the saddle point, we have
\begin{equation}
u_0=\frac{h\tl z}{2\sqrt{\wtl w}}.
%\label{9}
\end{equation}
Finally, for the intermediate range, the solution of Eq.~(7) can be written
in the approximate form
\begin{equation}
u_0\approx\biggl[\frac{\tl z^2h^2}{8\wtl w}-6+6\biggl[1+
\frac{\tl z^2h^2}{24\wtl w}+\frac{\tl\rho^2h^2}{12 \wtl w}+
\frac{\tl z^4h^4}{2304\wtl w^2}\biggr]^{1/2}\biggr]^{1/2}.
%\label{10}
\end{equation}
It is easy to see that this solution in the limits of weak ($h\ll1$) and
strong ($h\gg1$) fields respectively transforms into~(8) and~(9). In this
case, the evaluation of integral~(6) obtained by the saddle point
approximation is written as
\begin{equation}
I(\tl z,\tl\rho)\approx2\sqrt{\frac{\pi}h}\biggl[\frac{4\tl z^2}{u_0^2}
\sinh^2\biggl(\frac{u_0}2\biggr)+\frac{\tl\rho^2 u_0}2
\cth\biggl(\frac{u_0}2\biggr)\biggr]^{-1/2}e^{(u_0/2)-hg(u_0)}.
%\label{11}
\end{equation}

We now turn to studying the wave function in the limits of weak and
strong fields. We see in what follows that these concepts require
some more accurate definitions in the problem under consideration.
If $h\ll1$ and if $\tl\rho$ and $\tl z$ are not very large, then
solution~(8) reduces to
\begin{equation}
u_0=\frac{h\tl{r}}{2\sqrt{\wtl w}}\ll1,
%\label{12}
\end{equation}
where $\tl{r}=\sqrt{\tl z^2+\tl\rho^2}$ is the dimensionless radial
coordinate. In the limit under consideration, this expression in fact
determines the spherical type of symmetry of the electron cloud.

As already stated in the introduction, the behavior of the bound
energy level of a spinning particle located in a $\delta$-well in
the magnetic field was studied in~\cite {RodPhysRev}. It is
important that the obtained solution has stable asymptotic
expressions with respect to the spin variable in the limits of
weak and strong fields, i.e., the effective energy characteristic
$\wtl w$ is independent of the spin in both cases. Although the
computational technique used in~\cite {RodPhysRev} differs
slightly from techniques used for scalar particles in the previous
papers~\cite {Pop},~\cite {DemDr}, the results obtained in the
limits of weak and strong fields in~\cite {RodPhysRev} agree
completely. In particular, according to~\cite {RodPhysRev},we have
$\wtl w\approx1+h/2$ for small magnetic fields, whence it can be
directly seen that there is no dependence of $\wtl w$ on the
particle spin.

\begin{figure}[h]
\centering
\includegraphics{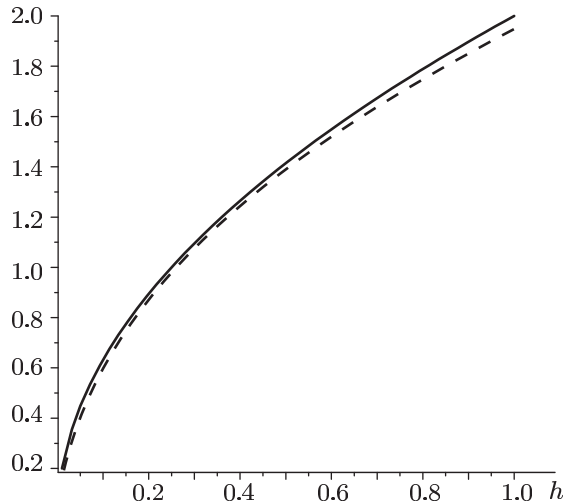}
\caption{The exact dependence of the function $\zeta(3/2,\wtl w/h)$ on the
magnetic field $h$ (the dashed curve) and the dependence determined by the
right-hand side of formula~(13) (the solid curve).}
%\label{fig2}
\end{figure}

Taking the foregoing into account, we use the estimate for the zeta function
\begin{equation}
\zeta\biggl(\frac32,\frac{\wtl w}{h}\biggr)\sim2h^{1/2},
%\label{13}
\end{equation}
which is applicable in the range $h\le1$. In particular, the
graphs shown in Fig.~4 demonstrate that such an asymptotic
expression is valid. Therefore, for the range under consideration
$h<1$ (and not very far from the $\delta$-well), we obtain the
expression
\begin{equation}
\psi(\vec{r}\,)\approx N\frac{2\sqrt{\pi}}{h\tl{r}}
\exp\biggl(-\frac i4h\tl x\tl y-\tl{r}\biggr)
%\label{14}
\end{equation}
for the spatial part of the wave function.

\begin{figure}[h]
\centering
\includegraphics{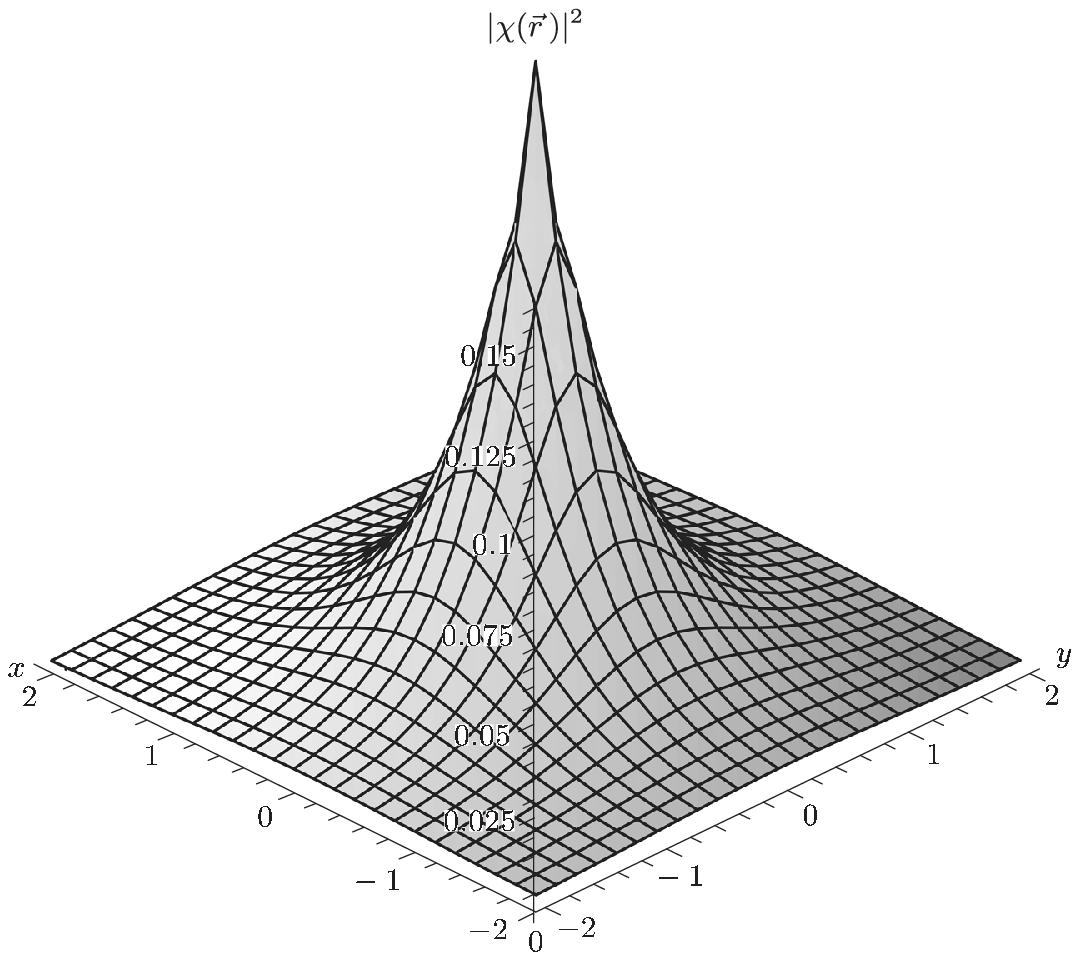}
\caption{The dependence of the dimensionless squared function $\xrs$ on
the transverse ($x$) and longitudinal ($z$) coordinates for fields $h\le1$.}
%\label{fig3}
\end{figure}

To represent the results graphically, it is convenient to use the function
$\chi(\vec{r}\,)=r\psi(\vec{r}\,)$ having no singularities at zero and
determining the spatial distribution of the probability density of the
electron cloud in the spherical system of coordinates:
$$
dW\sim 2\pi|\chi({\vec{r}}\,)|^2\sin(\theta)\,d\theta\,dr.
$$
Figure~5 shows the graph of $\xrs$. In this case, the de~Broglie
wavelength is assumed to be unity. We stress that this estimate
for the wave function was obtained in the vicinity of the
$\delta$-well in the weak field limit. But we note that the field
of the order of several tenths of the interatomic field is not
weak in the ordinary sense.

Far from the $\delta$-well, even in the weak field approximation,
the case changes cardinally. For $h\tl z\gg1$, from formula~(10), we
have
\begin{equation}
u_0=\frac{h\tl z}{2\sqrt{\wtl w}}\gg1
%\label{15}
\end{equation}
for the saddle point. As a result, the asymptotic expression for the spatial
part of the wave function becomes
\begin{equation}
\psi(\vec{r}\,)\approx N\sqrt{\frac{\pi}{1+h/2}}\exp\biggl(-\frac i4
h\tl x\tl y-\frac{h\tl\rho^2}8-\tl z\sqrt{1+\frac h2}\biggr).
%\label{16}
\end{equation}
This expression is important for a qualitative analysis. The
solution has an axial symmetry in this range. It is clear that the
value of the field determines the range where the spherical
symmetry typical of the bound s state transforms into the axial
symmetry inherent in the wave functions of particles in a purely
magnetic field. In addition, it can be seen that at large
distances from the $\delta$-well even in the weak-field case, only
the ground level of the effective energy makes the main
contribution, i.e., the level located at the smallest distance
from the continuum edge ($1+h/2$ in our adopted units). This
agrees completely with the conclusions in~\cite {DemDr},~\cite
{RodPhysRev}. In the case where the processes with free particles
in an external magnetic field are considered, it is usually
assumed that the weak field does not always exhibit its quantizing
character, yielding only small corrections to the cross sections
of the corresponding processes. In contrast, it is traditionally
assumed that only several minimum-energy levels contribute in the
strong-field case~\cite {Ter}.

In contrast to this, in the case of a bound state, any arbitrarily
weak field behaves as a strong field at large distances from the
center. In particular, it follows from~(16) that the magnetic
field compresses the electron cloud not only in the direction
transverse to the field (which can be expected) but also along the
field. But if the shift in the ground level of the effective
energy in the magnetic field is neglected, i.e., if it is assumed
that $\wtl w=1$, then such effects cannot be described. We note
that under these simplifying assumptions, solution~(16) agrees
with the basic function used in~\cite {Kh1}.

For a strong field, the case is more complicated. The reason is
obvious because the intermediate range in which the symmetry
transforms from the spherical into the axial one in this case
belongs to the range where the wave function differs significantly
from zero. Using the results in~\cite {RodPhysRev} on determining
the effective energy of the electron in strong fields ($h\gg1$),
$\wtl w\approx0.295h$, we consider a small neighborhood of the
$\delta$-center. For $\tl{r}\ll1/\sqrt{h}$, the position of the
saddle point is determined by expression~(12). For the wave
function, using relations~(4) and~(11), we obtain an analogue of
expression~(14) but with a different effective energy:
\begin{equation}
\psi(\vec{r}\,)\approx N\frac{2\sqrt{\pi}}{h\tl{r}}
\exp\biggl(-\frac i4h\tl x\tl y-\tl{r}\sqrt{0.295h}\biggr).
%\label{17}
\end{equation}
Therefore, supplementing our conclusions on the behavior of the
function at large distances in a weak field, we note that any
arbitrarily strong field behaves as a weak one in the direct
vicinity of the attractive $\delta$-center. The reason is that the
depth of the $\delta$-well is much greater than any finite shift
in the energy level in the magnetic field. The equation for the
bound level energy $w(h)$ follows precisely from the fact that the
character of this asymptotic expression is independent of the
external magnetic field~\cite {Pop},~\cite {DemDr},~\cite
{RodPhysRev}.

\begin{figure}[h]
\centering
\includegraphics{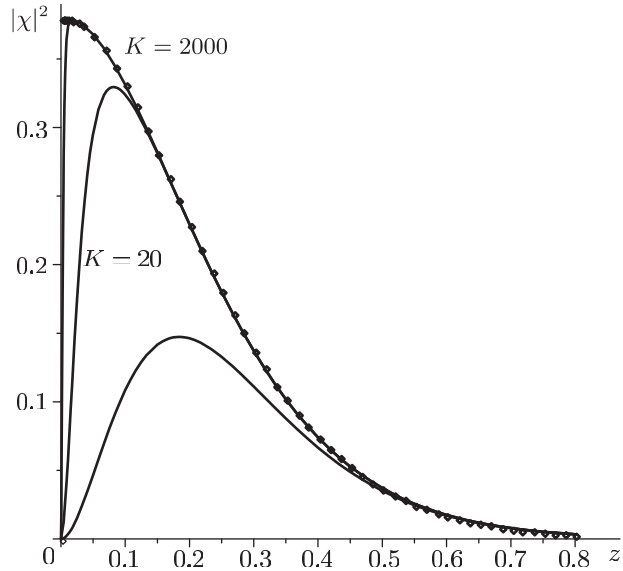}
\caption{The dependence of $\xrs$ on $z$ on the axis $\rho=0$ for
the field $h=100$. The points correspond to the calculation using
formula~(4), and the solid curves (successively from the bottom up)
determine the strong field approximations: the contribution of the
ground level~(18) and the contributions for $K=20$ and $K=2000$
calculated using formula~(20).}
%\label{fig4}
\end{figure}

The wave function becomes axially symmetric in the strong field at distances
$\tl z\gg1/\sqrt{h}$. For $h\gg1$ and fixed $\tl\rho\ne0$ and $\tl z\ne0$,
we obtain formula~(15). For the wave function, from relations~(4) and~(11),
we obtain an analogue of formula~(16) in the case of strong fields:
\begin{equation}
\psi(\vec{r}\,)\approx N\sqrt{\frac{\pi}{0.295h}}\exp\biggl(-\frac i4
h\tl x\tl y-\frac{h\tl\rho^2}8-\tl z\sqrt{0.295h}\biggr).
%\label{18}
\end{equation}
It is easy to see that also in this case, only the ground level of the
effective energy contributes.

Spherically symmetric estimate~(12) and axially symmetric
estimate~(15) of the maximum point agree in a neighborhood of the
straight line $\tl\rho=0$. It is interesting that the dependence
of the bound level energy on the magnetic field was first obtained
in~\cite {DemDr}, where the limit transition in the expression for
the wave function was used precisely along this line. Therefore,
to extend the applicability range of the obtained expression~(18)
to the range of small $z$, we can expand the denominator in
initial formula~(4) into a geometric series in terms of partial
effective levels,
\begin{equation}
{\wtl w}_k=\wtl w+kh,\quad k=0,1,2,\dots,
%\label{19}
\end{equation}
replacing the Landau levels in a purely magnetic field~\cite
{Lan}, and use estimate~(18) for each term. The wave function in
this case becomes
\begin{equation}
\psi(\vec{r}\,)\approx N\sqrt{\pi}\exp\biggl(-\frac i4h \tl x\tl y-
\frac{h\tl\rho^2}{8}\biggr)\sum_{k=0}^K
\frac{e^{-\tl z\sqrt{(0.295+k)h}}}{\sqrt{(0.295+k)h}}.
%\label{20}
\end{equation}
It is easy to verify that if the maximum number $K$ of levels taken
into account increases, then the applicability range for the
obtained formula can be extended to arbitrarily small values of the
coordinate~$z$.

In Fig.~6, the squared function $\xms$ calculated using exact
formula~(4) is compared with asymptotic expressions~(18) and~(20)
for the field $h=100$ and the axis $\rho =0$. The larger the
parameter $\sqrt{h}z$ is, the better the series in solution~(20)
converges (and the first term of series~(18) consequently
approaches the exact solution). This turns out to be important
because the probability current is maximum in this range.

\begin{figure}[h]
\centering
\includegraphics{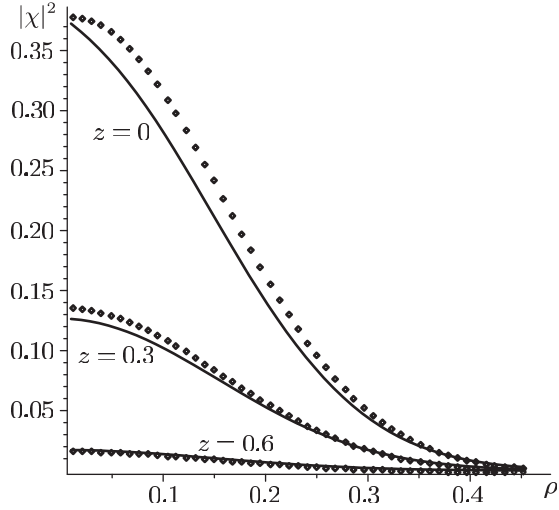}
\caption{The dependence of $\xms$ on $\rho$ for the field $h=100$ and
different $z$. The points correspond to the calculation using exact
formula~(4), and the solid curves correspond to strong field approximations
using formulas~(10) and~(11).}
%\label{fig5}
\end{figure}

Figure~7 shows the comparison of the expressions for $\xms$ in the
case of three different values of $z$ ($z=0,0.3,0.6$) and the
strong field $h=100$ calculated using exact formula~(4) and
approximate formula~(11) with~(10) taken into account.

We briefly summarize the results in this section. The solution of
the Pauli equation is spherically symmetric for a weak magnetic
field in the entire relevant range of coordinates. It is in fact
determined only by the character of the singularity in the
$\delta$-well; all effective partial levels $\wtl w_k$ contribute to
the formation of this singularity (see formula~(19)). The quantizing
character of the magnetic field (i.e., the tangibility of the
contribution of individual energy levels) is manifested only far
from the center in the region where the wave function is
exponentially suppressed. For the strong field, the character of the
solution symmetry changes in the relevant range of coordinates. The
relative contribution of individual levels for a fixed field
strength is mainly determined by the longitudinal coordinate (in the
magnetic field direction), although the electron cloud is mainly
compressed by the magnetic field in the transverse direction.

\section{{Probability currents of a particle bound by the
$\delta$-potential in a magnetic field}}
%\label{s3}

The general expression for the nonrelativistic probability current
of the spinor was given in the classic monograph~\cite {Lan}.
Using the usual tensor notation, we represent this current $J_k$
in the form
\begin{equation}
J_k=-i\frac{\hbar}{2m}[(\nabla_k\psi^*_{\alpha})\psi_{\alpha}-
\psi^*_{\alpha}\nabla_k\psi_{\alpha}]+\frac e{mc}A_k\psi^*_{\alpha}
\psi_{\alpha}-\frac{\mu c}{e|s|}\vep_{kpq}\nabla_p(\psi^*_{\alpha}
\widehat{\sigma}^q_{\alpha \beta}\psi_{\beta}),
%\label{21}
\end{equation}
where $k=1,2,3$ are the spatial coordinate indices, $\alpha=1,2$ are
the spin indices, $\nabla_k$ is the usual gradient operator, $A_k$
is the vector potential of the external magnetic field, $\mu$ is the
electron magnetic moment, $\vep_{kpq}$ is the unit totally
antisymmetric tensor, and $\widehat{\sigma}^q_{\alpha\beta}$ are the
binary sigma matrices. The complete normalized solution of the Pauli
equation can be represented in the form
$$
\psi_{\alpha}(\vec{r}\,)=\psi(\vec{r}\,)
\begin{pmatrix}1/2+s\\[1mm]1/2-s\end{pmatrix},
$$
where the spatial part $\psi(\vec{r}\,)$ is given by~(4). The first and
second terms in~(21) determine the gradient (orbital) current, and the last
term corresponds to the spin current.

Contracting spin indices with the explicit form of the sigma
matrices taken into account, we obtain the spatial components of the
total probability current
\begin{align*}
&J_x=-i\frac{\hbar}{2m}\biggl(\frac{\ptl\psi^*}{\ptl x}\psi-
\psi^*\frac{\ptl\psi}{\ptl x}\biggr)+\frac e{mc}A_x\psi^*\psi-
\frac{\mu c}e\frac s{|s|}\frac{\ptl}{\ptl y}(\psi^*\psi),
\\[2mm]
&J_y=-i\frac{\hbar}{2m}\biggl(\frac{\ptl\psi^*}{\ptl y}\psi-
\psi^*\frac{\ptl\psi}{\ptl y}\biggr)+\frac e{mc}A_y\psi^*\psi+
\frac{\mu c}e\frac s{|s|}\frac{\ptl}{\ptl x}(\psi^*\psi),
\\[2mm]
&J_z=-i\frac{\hbar}{2m} \biggl(\frac{\ptl\psi^*}{\ptl z}\psi-
\psi^*\frac{\ptl\psi}{\ptl z}\biggr)+\frac e{mc}A_z\psi^*\psi.
\end{align*}

We must fix the gauge of the vector potential for further calculations.
Choosing it as  $A_x=-Hy$ and $A_y=A_z=0$ and passing to the dimensionless
coordinates, we obtain
\begin{equation}
\vec{J}=\biggl(\frac{|W_0|}{2m}\biggr)^{1/2}\frac h2\psi^*\psi(\vec{j}\tl x-
\vec{i}\tl y)+\frac{(2m|W_0|)^{1/2}}{2\hbar}\frac{\mu c}e\frac s{|s|}
\psi^*\wtl{\psi}(\vec{j}\tl x-\vec{i}\tl y),
%\label{22}
\end{equation}
where $\vec{i}$ and $\vec{j}$ are the unit vectors of the Cartesian system
and the function $\wtl{\psi}(\vec{r}\,)$ has the form
$$
\wtl{\psi}(\vec{r}\,)=Ne^{-ih\tl x\tl y/4}
\int_0^{\infty}\frac{d\tau}{\tau^{1/2}}\frac{1+e^{-h\tau}}{[1-e^{-h\tau}]^2}
\exp\biggl[-\frac{\tl z^2}{4\tau}-\frac h8\tl\rho^2\cth
\biggl(\frac{h\tau}2\biggr)-\wtl w\tau\biggr].
$$
Assuming that the electron magnetic moment is equal to the Bohr magneton, we
obtain
\begin{equation}
\vec{J}=\biggl(\frac{|W_0|}{2m}\biggr)^{1/2}
\frac h2\psi^*(\psi+2s\wtl{\psi})(\vec{j}\tl x-\vec{i}\tl y).
%\label{23}
\end{equation}

Using exact expression~(4) for the spatial part of the wave
function, we obtain the formula for the probability current in the
case of an arbitrary spin orientation and field strength:
\begin{align}
{\vec{J}}={}& Mh^{7/2}\zeta^{-1}\biggl(\frac32,\frac{\wtl w}h\biggr)
(\vec{j}\tl x-\vec{i}\tl y)\times{}
\nonumber
\\[2mm]
&{}\times\int_0^{\infty}\frac{d\tau}{\tau^{1/2}}
\frac{\exp\bigl[-\tl z^2/(4\tau)-h\tl\rho^2\cth(h\tau/2)/8-
\wtl w\tau \bigr]}{1-e^{-h\tau}}\times{}
\nonumber
\\[2mm]
&{}\times\int_0^{\infty}\frac{d\tau}{\tau^{1/2}}
\frac{\exp[-\tl z^2/(4\tau)-h\tl\rho^2 \cth(h\tau/2)/8-\wtl w\tau]}
{1-e^{-h\tau}}\times{}
\nonumber
\\[2mm]
&{}\times\biggl[1+2s\cdot\cth\biggl(\frac{h\tau}2\biggr)\biggr],
%\label{24}
\end{align}
where $M=m|W_0|^2/4\pi^2\hbar^3$ is a dimension factor independent
of the magnetic field.

We obtain the estimates for the current in the weak-field approximation. In
this case, we have
$$
\wtl{\psi}(\vec{r}\,)\approx N\cdot\frac{8\sqrt{\pi}}{h^2\tl{r}^3}(1+\tl{r})
\exp\biggl(-\frac i4h\tl x\tl y-\tl{r}\biggr).
$$
Also taking~(14) and~(23) into account, we write the expression for the
probability current in the weak magnetic field:
\begin{equation}
\vec{J}=M\zeta^{-1}\biggl(\frac32,\frac1h+\frac12\biggr)
\frac{4\pi h^{3/2}}{\tl{r}^2}e^{-2\tl{r}}\biggl[1+2s\cdot
\frac{4(1+\tl{r})}{h\tl{r}^2}\biggr](\vec{j}\tl x-\vec{i}\tl y).
%\label{25}
\end{equation}

\begin{figure}[h]
\centering
\includegraphics[width=116mm]{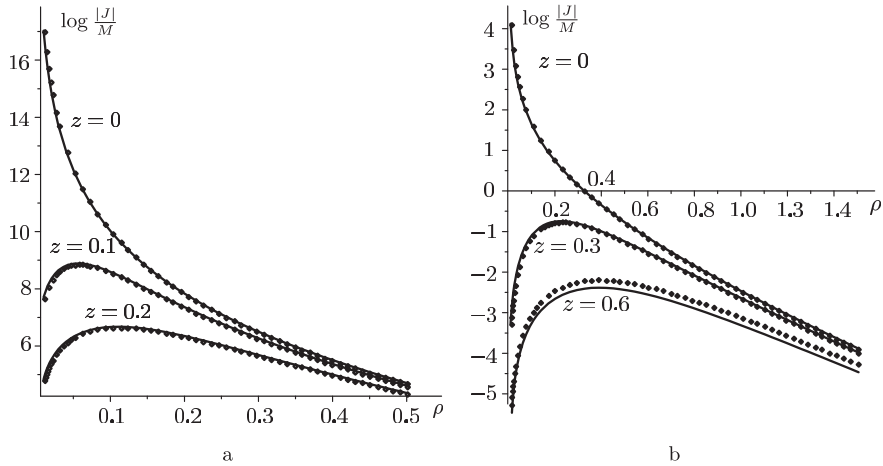}
\caption{The dependencies of $\ln\big(|J|/M\big)$ on the
transverse coordinate in the planes $z=\text{const}$ for the weak
field $h=0.1$ in the cases of (a)~the particle with spin directed
opposite the field and (b)~the scalar particle. The points
correspond to the calculation using exact formula~(24), and the
solid curves correspond to weak-field approximation~(25).}
%\label{fig6}
\end{figure}

\begin{figure}[h]
\centering
\includegraphics[width=116mm]{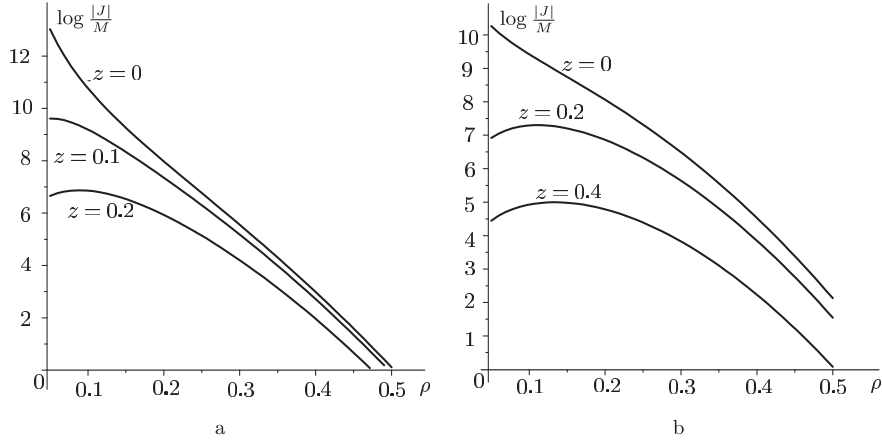}
\caption{The dependencies of $\ln\big(|J|/M\big)$ on the
transverse coordinate calculated using exact formula~(24) in the
planes $z=\text{const}$ for the strong field $h=100$ in the cases
of (a)~the particle with the spin directed opposite the field and
(b)~the scalar particle.}
%\label{fig7}
\end{figure}

Figure~8a shows the dependencies of the logarithm of $|J|/M$ on
the transverse coordinate $\rho$ in different planes
$z=\text{const}$ for the field $h=0.1$; they are calculated using
exact and approximate formulas~(24) and~(25) for $s=-1/2$, i.e.,
for the particle with the spin oriented opposite the field. In the
case of the particle with the spin along the field, only the
orientation of the vector $\vec{J}$ changes, i.e., the direction
of the particle rotation changes. This occurs because the second
term in~(25) in the vicinity of the $\delta$-well turns out to be
much greater than the first term. In other words, the orbital
probability current in this case is negligibly small compared with
the spin one. They become comparable only at a distance
$\tl{r}\sim 1/h$ from the center, where the current in the weak
field~($h\ll1$) is exponentially suppressed. This conclusion is
confirmed by comparing Fig.~8a with Fig.~8b, where the same
quantity $\ln\big(|J|/M\big)$ is constructed for the scalar
particle. In this case, the spatial part of the wave function in
the weak field (see formula~(14)) is independent of the spin. The
obtained result is quite expectable if we recall that in the case
of a weak field, the $\delta$-well itself can bind a charged
particle only if it is in the s state.

The dependencies of $\ln(|J|/M)$ on the transverse coordinate in
the planes with different $z$ for the strong magnetic field
$h=100$ demonstrating the spatial distribution of the probability
currents are shown in Fig.~9a for the electron with the spin
directed opposite the field and in Fig.~9b for the scalar
particle. Exact formula~(24) is used in the calculation.

\begin{figure}[h]
\centering
\includegraphics{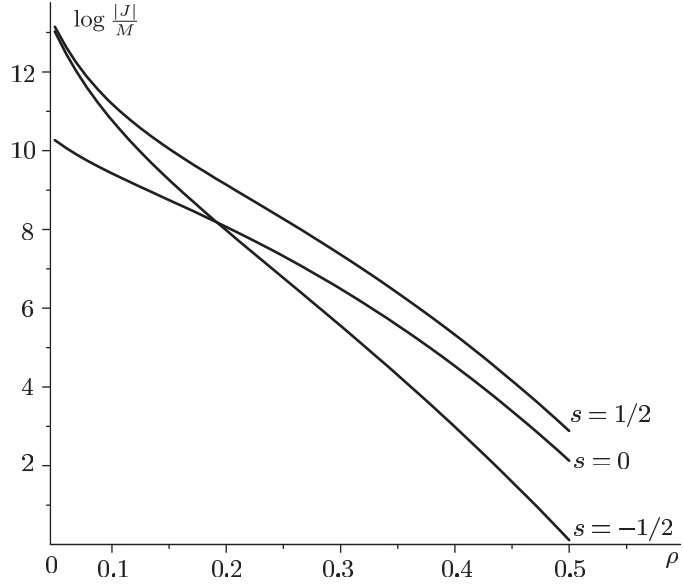}
\caption{The dependence of $\ln\big(|J|/M\big)$ on the transverse coordinate
obtained using exact formula~(24) in the case of the plane $z=0$ and the
strong field $h=100$ for different spin orientations (along the field and
opposite the field) and the zero spin.}
%\label{fig8}
\end{figure}

Figure~10 shows the comparison of the behavior of $\ln(|J|/M)$ for
the scalar particle with that for the electron with its different
spin orientations in the case of the strong field and the plane
$z=0$. In particular, it can be seen from Fig.~8 that the orbital
and spin currents in the strong field become comparable in the
vicinity of the $\delta$-center. As should be expected, in the
case of the electron with the spin oriented opposite the magnetic
field, the current decreases more rapidly as the distance from the
center increases, i.e., this particle motion is localized. The
range where the orbital and spin currents turn out to be
comparable is probably most interesting for studies.

\subsection*{Acknowledgments}
This work was supported in part by the Program for Supporting
Leading Scientific Schools (Grant Nos.~NSh-8265.2010.1, G.~A.~K.)
and the Russian Foundation for Basic Research (Grant
No.~09-02-00725\_a).

\end{document}